\documentclass{article}

\usepackage{amssymb}

\usepackage{graphicx}
\usepackage{epsfig}

\begin{document}

\title{An Electrostatic Lens to Reduce Parallax in Banana Gas Detectors.}

\author{P. Van Esch \\
J-F. Clergeau \\
K. Medjoubi}

\maketitle

\begin{abstract}

Cylindrical "banana" gas detectors are often used in fixed-target
experiments, because they are free of parallax effects in the
equatorial plane. However, there is a growing demand to increase
the height of these detectors in order to be more efficient or to
cover more solid angle, and hence a parallax effect starts to
limit the resolution in that direction. In this paper we propose a
hardware correction for this problem which reduces the parallax
error thanks to an applied potential on the front window that
makes the electrostatic field lines radially pointing to the
interaction point at the entrance window.  A detailed analytical
analysis of the solution is also presented.

\end{abstract}

\section{Introduction}
So-called "banana" detectors are large-area, one-dimensional or
2-dimensional detectors that have a concave cylindrical detection
surface.  The axis of the cylinder passes through the sample
position, and the detector is used to measure the \(\phi \) angle
of scattered radiation from the sample in the case of a
one-dimensional detector, or to measure \(\phi\) and \(z\) in a
two-dimensional detector.  The advantage of also measuring \(z\)
is that, in the case of anisotropic samples, one can extract
\(\theta\) information, or in the case of isotropic samples, one
can correct for the projection of the cone with opening angle
\(\phi\) projected onto a cylindrical surface.  Increasing the
\(z\)-aperture of the detector increases of course the counting
efficiency in the case of isotropic samples, and the spherical
angle covered in the case of anisotropic samples.

If the detecting volume has a non-negligible thickness, as is
often the case for gas detectors, ideally, along the radial line
(pointing towards the sample) any detection should give rise to an
equivalent "position" detection, which is actually a radial
direction detection.  In gas detectors, this is a priori not so:
the drift field in the detection volume is usually perpendicular
to the cylindrical wall, so there will be a dependence of the
measured \(z\)-value on the exact detection point along the
particle path (a radial line). In the case of neutral particle
detection, the distance inside the detection volume of the
detection point along the path is usually a random variable with
an exponential distribution. A very narrow beam along a
well-defined angle $\theta$ will give rise to a spread in
\(z\)-values. This loss in resolution, together with a change of
center of gravity of the impact distribution, is called the
parallax effect.

The way to avoid this parallax error is to have the electrostatic
drift field be radial until the final signal generation.  Of
course, within a cylindrical geometry, this will not be possible
if the final signal generating (gas amplification) surface is
equipotential: there the field lines will have to be perpendicular
to the cylindrical surface.  But one can try to establish at least
in the first part of the drift volume a more or less radial
electric drift field.  A review of parallax correction techniques
is given in \cite{articlecharpak1982}.  A straightforward method
is to use a curved entrance window at constant potential, as done
by the authors of \cite{articlechernenko1995a} and
\cite{articlechernenko1995b}. The problem with this approach is of
course that the conversion gap changes considerably over the
\(z\)-range and it is application-dependent if this is acceptable
or not.   We have been inspired more by the technique described in
\cite{brevetcomparat1988} and \cite{articlebrookhaven1997}, to
turn the parallel drift field into a radial drift field, at least
in the first part of the drift volume.  Given the structure of the
`banana' detector, we only wish to deflect the field in the
\(z\)-direction.  It turns out that, provided one can make a few
approximations, this problem leads to an analytic solution for the
potential to be applied at the entrance window.

\section{Analytical solution.}

We are looking for a potential function \(V(z)\), to be applied at
the entrance window, such that the E-field is pointing to the
sample position at the entrance window.  We assume the conversion
gap \(d\) to be much smaller than the sample distance \(R\).  As
such we make the following approximation: we assume that the
normal component of the electric field \(E_n\) is independent of
the coordinate \(\rho\) (measured perpendicularly on the cylinder
surface inward the detection volume) between the entrance window
and the detection plane (supposed to be at an equivalent constant
potential \(V_c\)). Note that we don't assume that the normal
component is independent of \(z\). This is a priori justified by
the smallness of the gap compared to all other potential
variations which should be of the order of the sample distance.
This results in our first equation:
\begin{equation}
\label{eq:normalfield}
E_n(z) = \frac{V(z)-V_c}{d}
\end{equation}
Next, we want the E-field to be radial at the entrance window,
leading to our second equation:
\begin{equation}
\label{eq:fieldradial}
\frac{E_t(z)}{E_n(z)}=\frac{z}{R}
\end{equation}
In this equation, \(E_t(z)\) is the tangent component of the
E-field at the entrance window.  It is determined by the potential
gradient:
\begin{equation}
\label{eq:tangentfield}
E_t(z) = -\frac{d V}{dz}
\end{equation}
Note that equations \ref{eq:fieldradial} and \ref{eq:tangentfield}
are exact; our only approximation is equation \ref{eq:normalfield}
which we justified, and which we will verify later. The three
equations give rise to a linear differential equation in \(V(z)\).
Solving this equation with the boundary condition that at height
\(h\), the potential \(V(z)\) has to vanish (the cage is supposed
to be at ground potential), we find:
\begin{equation}
\label{eq:solutiondifeq} V(z) = V_c\left( 1 - e^{\frac{h^2-z^2}{2
R d}}\right)
\end{equation}
which is nothing else but a Gaussian profile. To apply such a
profile, a suitable sheet (Kapton, for instance) with straight
metallic strips, connected by a resistive voltage divider, can be
applied. If the strips are regularly spaced at a distance
\(\Delta\), then of course the resistor values should be chosen
as:
\begin{equation}
r_k = r \left[V(k \Delta) - V\left( (k-1) \Delta\right)\right]
\end{equation}
where \(r_k\) is the resistor linking the \(k\)-th strip to the
previous one, starting from the strip in the middle (at \(z=0\)).
\(r\) is an arbitrary overall scale factor for the resistors,
which will determine the total resistance of the voltage divider.

\section{Field and drift line calculation.}

We apply the above potential in a specific example, namely the
geometry for the future D19 banana thermal neutron detector at the
ILL: an active height of 40cm ($h=20cm$), a conversion gap of
$d=2.6cm$ and a sample distance of $R = 70cm$.  There are 4cm of
extra space on top and at the bottom in the \(z\) direction: the
entrance window will continue to be at ground potential over these
4cm on each side, as well as the top cover and the bottom cover.
The detection "plane" actually consists of a layer of cathode
wires at 26 mm from the entrance window, and a layer of anode
wires at 30 mm from the entrance window.  We will consider the
cathode wire plane to be an equipotential surface, and with the
applied potentials on the wires, this comes down to an equivalent
potential of $V_c=600V$.

\begin{figure}
  \includegraphics[width=8cm]{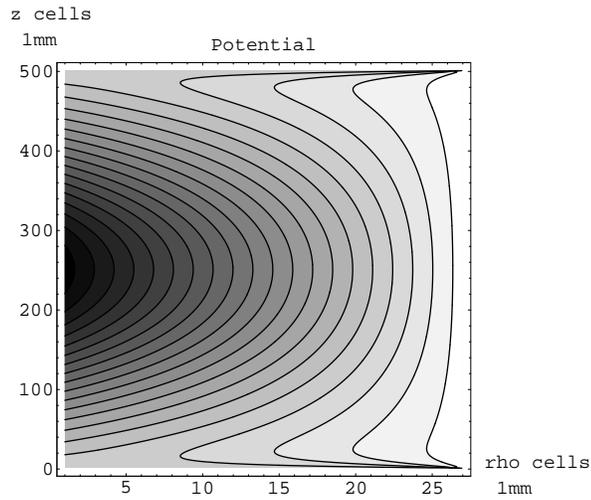}\\
  \caption{The potential, solved by the relaxation method.}\label{fig:potential}
\end{figure}

\begin{figure}
  \includegraphics[width=8cm]{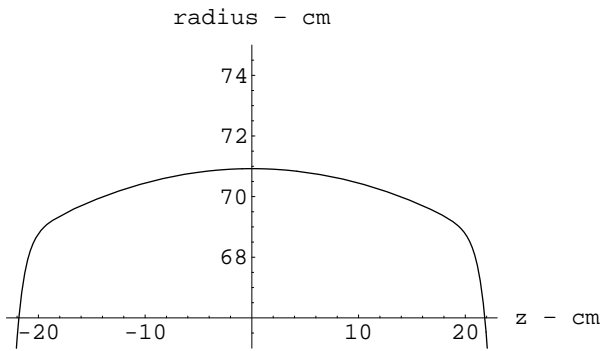}\\
  \caption{The intersection point of the tangent to the drift line at the entrance
  window and the central axis.  Ideally, this should be equal to the sample distance of 70cm
  everywhere.}\label{fig:radiuscheck}
\end{figure}

Using a 1mm \(\times\) 1mm grid of points in the rectangular drift
volume, and applying the relaxation method as explained on p47 of
\cite{jackson}, iterating 3000 times yields an accuracy of better
than \(30mV\).  The solution for the potential can be seen in
figure \ref{fig:potential}.  One can check that the original goal
set forth, namely that the electric field at the level of the
entrance window points towards the sample position, is satisfied
to a high degree, as shown in figure \ref{fig:radiuscheck} , which
justifies our approximation after the fact.

This solution comes close to the following analytical expression:
\begin{equation}
V_{an}(z,\rho) = V_c \frac{ d + (\rho-d) e^{\frac{h^2-z^2}{2 R d}}
}{d}
\end{equation}
On the entrance window and on the detection surface, the
expression is exact, and we have a linear interpolation
perpendicular to the entrance window. This is another way of
stating our approximation as given in equation
\ref{eq:normalfield}. Clearly, this analytical expression is not a
harmonic function and hence cannot be the true solution, but when
we take the difference between the numerical values given by this
expression and the values found by the relaxation technique, the
difference is less than $7.0V$ (maximum error in the middle of the
drift volume), which is on the 1\% level.  The advantage of such a
simple expression over numerical results is that solving for the
drift lines is possible analytically. Indeed, working out the
differential equation for the drift line $z(\rho)$ as a function
of the condition $z(\rho_0) = z_0$ gives us the solution:
\begin{equation}
z(\rho) = z_0 \exp \left(
{\frac{(2d-\rho-\rho_0)(\rho-\rho_0)}{2dR} }\right)
\label{eq:driftline}
\end{equation}
According to this curve, a neutron that converts in position
$(z_1,\rho_1)$ is projected onto the $z$-value:
\begin{equation}
z_{pr}(z_1,\rho_1) = z_1 \exp
\left(\frac{(d-\rho_1)^2}{2dR}\right)
\end{equation}
Clearly without the electrostatic lens, the exponential factor is
absent.

\section{Projected image of a ray.}

We consider an incident beam from the sample position under an
angle $\theta$. After having travelled a distance $s$ in the
conversion gap, the particle is at position $(\rho_1 = \cos \theta
s, z_1 = R \tan \theta + \sin \theta s)$. The ray will thus give
rise to hits which are confined between $R e^{ \frac{d}{2R} }\tan
\theta \simeq (R+\frac{d}{2}) \tan \theta$ and $(R+d) \tan
\theta$. Without lens, this ray would give rise to hits between $R
\tan \theta$ and $(R+d) \tan \theta$, which means that the carrier
of the hit distribution with lens is divided by 2 as compared to
without a lens. But in order to work out more accurately the
improvement upon the parallax error, we need to work out the
projected conversion density.

If we consider an absorption (conversion) constant $\mu$, the
projected density $\xi_{\theta}(z)$ of a narrow beam emanating
from the sample under an angle $\theta$ can be found by working
out:
\begin{equation}
\xi_{\theta}(z) = \int_{s=0}^{s_{max}}\mu e^{-\mu s} \delta \left[
z - (R \tan \theta + s \sin \theta)  \exp\left(\frac{(d - s \cos
\theta )^2}{2 d R}\right) \right] ds
\end{equation}
with $s_{max} = d /\cos \theta$. In order to solve this integral,
we need to know $s_0(z,\theta)$, the solution to the equation:
\begin{equation}
(R \tan \theta + s_0 \sin \theta)  \exp\left(\frac{(d - s_0 \cos
\theta )^2}{2 d R}\right) = z
\end{equation}
As such, this equation cannot be solved analytically.  However,
noting that $d/R \ll 1$, we can expand the exponential term, and
limit ourselves to first-order contributions in $d/R$. This
approximate solution is given as follows:
\begin{equation}
s_0(z,\theta)  \simeq {\sqrt{d}}{\sqrt{\csc \theta}}{\sqrt{\sec
\theta}}
  {\sqrt{2z - \left( d + 2R \right) \tan \theta}}
\end{equation}

If $s_0(z,\theta)$ is between 0 and $s_{max}$, which comes down to
requiring that $z$ is within the limits of the carrier of
$\xi_{\theta}$, we finally find for the projected density of a ray
under the effect of the lens:
\begin{equation}
\xi_{\theta}(z) =     \frac{2dR\csc \theta \mu e^{-\mu
s_0(z,\theta)}}
  {d^2 + 2\left( -2d + R \right) {\sqrt{d\cot \theta}}\,
     {\sqrt{2z - \left( d + 2R \right) \tan \theta}}}
\end{equation}

Without the lens, the projected density is given by:
\begin{equation}
\xi_{\theta}^0(z) = \int_{s=0}^{s_{max}}\mu e^{-\mu s} \delta
\left[ z - (R \tan \theta + s \sin \theta) \right] ds
\end{equation}
which has as a solution:
\begin{equation}
\xi_{\theta}^0(z) = \mu \exp \left( -\mu \left( \frac{z}{\sin
\theta} - \frac{R}{\cos \theta}\right) \right) \frac{1}{\sin
\theta}
\end{equation}
where it is understood that $z$ is within the boundaries of the
carrier of $\xi^0_{\theta}$.

\section{Resolution improvement: an example.}

\begin{figure}
  \includegraphics[width=8cm]{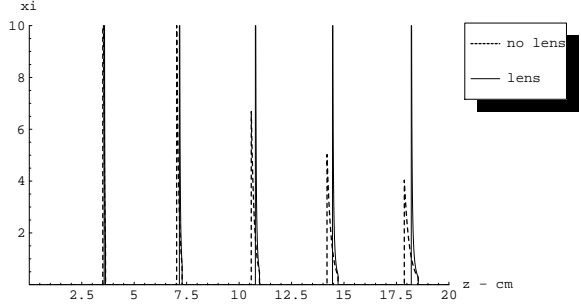}\\
  \caption{The projected densities, with and without lens, for $\mu = 1 cm^{-1}$.
  The represented beams have angles 0.05,0.1,0.15,0.20 and 0.25 rad.}
  \label{fig:projectiondensities}
\end{figure}

We will now take as an example the case $\mu = 1 cm^{-1}$, and the
geometry of the D19 prototype detector described earlier. If we
compare the projected densities for different beams, we obtain
profiles as shown in figure \ref{fig:projectiondensities}.

\begin{figure}
  \includegraphics[width=8cm]{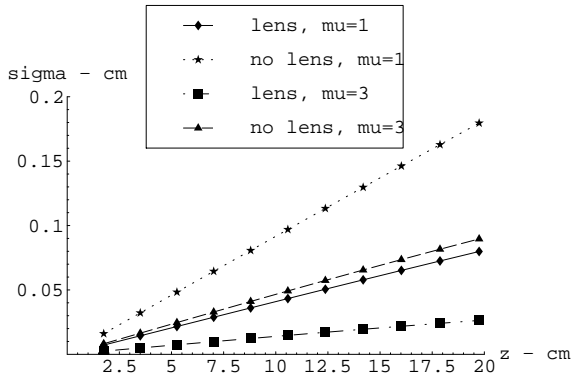}\\
  \caption{Resolution (standard deviation of projected density) as a function of
  the impact position at the front window, for the cases $\mu = 1 cm^{-1}$ and
  $\mu = 3 cm^{-1}$, at a fixed impact angle $\theta = 0.275$.}\label{fig:resolutions}
\end{figure}

\begin{figure}
  \includegraphics[width=8cm]{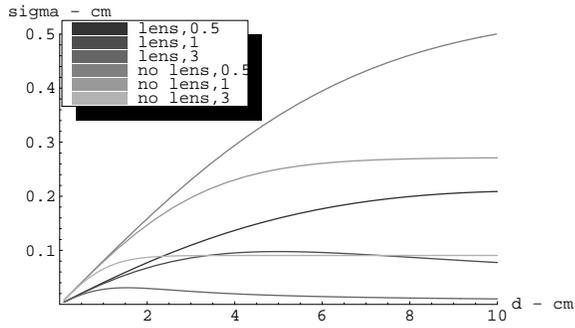}\\
  \caption{Resolution (standard deviation of projected density) as a function of
  the thickness of the conversion volume, for the cases $\mu = 0.5 cm^{-1}$, $\mu = 1 cm^{-1}$ and
  $\mu = 3 cm^{-1}$.}
  \label{fig:resolutionsthickness}
\end{figure}

We calculate the standard deviations of these distributions by
numerical integration, and also of the distributions obtained with
$\mu = 3.0 cm^{-1}$. We obtain the result shown in figure
\ref{fig:resolutions}. The resolutions are improved using a lens
by a factor slightly better than 2.2 for the case $\mu = 1
cm^{-1}$ and a factor slightly better than 3.3 for $\mu = 3
cm^{-1}$.  If we double the conversion gap (from 2.6 mm to 5.2
mm), and we do the calculation again, we find that for $\mu = 1
cm^{-1}$ the improvement is a factor of about 2.6 and for $\mu = 3
cm^{-1}$ the improvement is even a factor of about 5.5.  This
means that the relative improvement of the resolution error due to
parallax increases as well when $d$ as well as when $\mu$ become
large. However, we should pay attention to the absolute
resolutions as a function of thickness.  In figure
\ref{fig:resolutionsthickness}, we can observe that the resolution
without lens has a saturating behavior as a function of conversion
gap thickness, while the behavior with lens is better, but more
involved, in that the resolution reaches a maximum, and then
decreases when the conversion gap gets bigger.

\begin{figure}
  \includegraphics[width=8cm]{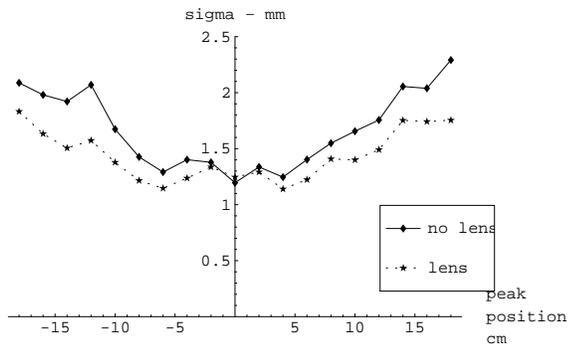}\\
  \caption{The measured resolution (standard deviations of a gaussian fit
  in mm) as a function
  of impact position.}\label{fig:experimental}
\end{figure}

Experimentally, the prototype has been filled with 5.5 bar of
He-3.  Neutrons with a wavelength of 2.5 Angstrom are used, which
comes down to a conversion factor $\mu = 0.99 cm^{-1}$.   A Cd
mask with 2 mm wide slits was put in front of the entrance window,
and a piece of plexiglass irradiated at the sample position
diffused the narrow neutron beam in a more or less radially
uniform way.  Activating or not, the electrostatic lens, we obtain
the resolutions, obtained by fitting a gaussian curve with offset
to the slit images using a least-squares algorithm, as shown in
figure \ref{fig:experimental}. The resolution finds its origin in
several different effects (intrinsic gas resolution, electronic
noise, quantization noise,...), but the deterioration due to
parallax is clearly visible, and is improved upon by the lens
action.

\section{Conclusion}

A solution has been presented to improve upon the parallax effect
in large-aperture banana detectors, using an electrostatic lens.
An approximate analytical expression of the expected image of a
ray is deduced. Calculations indicate that the relative
improvement upon the parallax error with this method becomes
stronger when both $\mu$ and $d$ are large. The absolute
resolution, with lens, has a more complicated behavior as a
function of $d$. An experimental verification of the improvement,
using the prototype of the new D19 thermal neutron detector at the
ILL, indicates qualitatively that the technique works in practice.


\begin{thebibliography}{10}
\bibitem{articlecharpak1982}
G. Charpak, Nucl. Instr. Meth. 201 (1982), 181-192.

\bibitem{articlechernenko1995a}
Yu.V. Zanevsky, S.P. Chernenko et {\it al.}, Nucl. Instr. Meth. A
367 (1995) 76-78

\bibitem{articlechernenko1995b}
Yu.V. Zanevsky, S.P. Chernenko et {\it al.}, Nucl. Phys. B 44
(1995) 406-408

\bibitem{brevetcomparat1988}
V. Comparat et al, French patent n° 2 630 829 (1988).

\bibitem{articlebrookhaven1997}
P. Rehak, G.C. Smith and B. Yu,  IEEE Trans. Nucl. Sci., vol 44,
no. 3 (1997) 651-655.

\bibitem{jackson}
J. D. Jackson, Classical Electrodynamics, third edition,
\copyright John Wiley 1999.
\end{thebibliography}
\end{document}